\documentclass{article}

\usepackage[final]{neurips_2025}

\usepackage[utf8]{inputenc} 
\usepackage[T1]{fontenc}    
\usepackage{hyperref}       
\usepackage{url}            
\usepackage{booktabs}       
\usepackage{amsfonts}       
\usepackage{nicefrac}       
\usepackage{microtype}      
\usepackage{xcolor}         
\usepackage{amsmath}
\usepackage{graphicx} 
\usepackage{subcaption} 
\usepackage{multirow,siunitx}
\usepackage{natbib}
\usepackage{makecell}
\usepackage{algorithm}
\usepackage{algpseudocode}
\usepackage{setspace}
\usepackage{amssymb}
\usepackage{wrapfig}
\usepackage{pifont}
\newcommand{\xmark}{\ding{55}}

\title{Parallax: Runtime Parallelization for Operator Fallbacks in Heterogeneous Edge Systems}

\author{%
  Chong Tang \\
  University of Southampton \\
  UCL AI Centre \\
  Southampton, United Kingdom \\
  \texttt{chong.tang@soton.ac.uk} \\
  \And
  Hao Dai \\
  University College London \\
  UCL AI Centre \\
  London, United Kingdom \\
  \texttt{hao.dai@ucl.ac.uk} \\
  \And
  Jagmohan Chauhan \\
  University College London \\
  UCL AI Centre \\
  London, United Kingdom \\
  \texttt{jagmohan.chauhan@ucl.ac.uk} \\
}

\begin{document}

\maketitle

\begin{abstract}
The growing demand for real-time DNN applications on edge devices necessitates faster inference of increasingly complex models. Although many devices include specialized accelerators (e.g., mobile GPUs), dynamic control-flow operators and unsupported kernels often fall back to CPU execution. Existing frameworks handle these fallbacks poorly, leaving CPU cores idle and causing high latency and memory spikes. We introduce Parallax, a framework that accelerates mobile DNN inference without model refactoring or custom operator implementations. Parallax first partitions the computation DAG to expose parallelism, then employs branch-aware memory management with dedicated arenas and buffer reuse to reduce runtime footprint. An adaptive scheduler executes branches according to device memory constraints, meanwhile, fine-grained subgraph control enables heterogeneous inference of dynamic models. By evaluating on five representative DNNs across three different mobile devices, Parallax achieves up to 46\% latency reduction, maintains controlled memory overhead (26.5\% on average), and delivers up to 30\% energy savings compared with state-of-the-art frameworks, offering improvements aligned with the responsiveness demands of real-time mobile inference.
\end{abstract}

\section{Introduction}
Efficient on-device inference of deep neural networks (DNNs) has become an integral part of modern mobile computing~\citep{xu2025fast,cai2019once}. As mobile applications shift from conventional recognition tasks to more sophisticated workloads, they increasingly contend with dynamic tensor shapes and control flow~\citep{shen2021nimble}. For example, object detectors predict a variable number of bounding boxes, and automatic speech recognition (ASR) employs beam search to dynamically adjust outputs. Meanwhile, continued innovation in DNN architectures introduces new operators that may lack universal support across hardware backends~\citep{ghodrati2024tandem}. 

State-of-the-art (SOTA) mobile inference frameworks, such as ONNXRuntime (ORT)~\citep{bai2019}, ExecuTorch~\citep{executorch2024}, and TensorFlow Lite (TFLite)~\citep{david2021tensorflow}, provide hardware delegation (GPUs, TPUs, NPUs) and compiler optimizations like operator fusion. However, they assume static graphs and predefined operators. When faced with dynamic or unsupported operations, frameworks fall back to CPU execution. TFLite often offloads the entire graph~\citep{niu2024sod2}, while ORT and ExecuTorch may reject or partially offload graphs with control flow~\citep{shen2021nimble}, leading to extra data transfers, memory copies and synchronization stalls~\citep{zhang2022comprehensive}. Recent solutions handle inference overhead from various perspective. ASPEN~\citep{park2023aspen} improves CPU-only operator-level parallelism by converting models into fine-grained tile graphs for dynamic scheduling. NN-Stretch~\citep{wei2023nn} and CoDL~\citep{jia2022codl} accelerate static graphs by scheduling different model segments to run concurrently on heterogeneous processors. SoD$^2$~\citep{niu2024sod2} addresses dynamic shapes by generating fixed-shape kernel variants offline based on operator dependencies and selecting the appropriate version at runtime. MikPoly~\citep{yu2024optimizing} follows a similar approach by compiling shape-specialized micro-kernels and linking only those matching observed tensor sizes. On the hardware side, Tandem Processor~\citep{ghodrati2024tandem} integrates a small programmable core alongside the main accelerator, allowing unsupported or irregular operators to run on-chip and avoid costly fallbacks to the host CPU.

While these methods offer promising improvements, they have practical limitations. Static code generation and micro-kernel libraries require continuous maintenance, as evolving models and operators can invalidate precompiled kernels~\citep{niu2024sod2,yu2024optimizing}. Hardware-centric solutions depend on new silicon and remain incompatible with most deployed mobile devices~\citep{ghodrati2024tandem}. Many other approaches like NN-Stretch require model refactoring, custom operator maintenance, or constrain model flexibility to preserve accelerator compatibility~\citep{wei2023nn,jia2022codl}. Thus, a clear need persists for an approach that utilizes readily available on-device resources to offset performance penalties from dynamic operations and unsupported operators.

To meet this need, we present Parallax, a scheduling framework that accelerates mobile DNN inference without model changes. Parallax non-invasively traverses the computation Directed Acyclic Graph (DAG), partitions it to expose parallelism, and assigns dedicated memory arenas per branch using efficient intra- and inter-branch buffer sharing. It then performs static shape inference and linear liveness scanning to estimate each branch’s peak memory. Branches are scheduled in parallel if within available system RAM (with safety margin), otherwise sequentially to prevent out-of-memory (OOM) issues. We evaluate Parallax on five DNNs across three mobile devices. Compared to existing solutions (Table~\ref{tab:comparison}), Parallax shows unique advantages. Furthermore, results demonstrate that Parallax reduces inference latency by up to 46\%, maintains controlled memory overhead (26.5\% average increase), and delivers up to 30\% energy savings, all without requiring model refactoring or custom operator implementations. Our key contributions are:
\begin{table}[t]
  \centering
  \caption{Parallax uniquely handles dynamic operators, avoids model/kernel modifications, supports heterogeneous inference and accelerates parallel CPU fallback execution.}
  \label{tab:comparison}
  \small
  \scalebox{0.9}{
  \begin{tabular}{lcccc}
    \toprule
    \textbf{Method} & \makecell{\textbf{Handles} \\ \textbf{Dynamic Ops}} & \makecell{\textbf{No Model Refactoring} \\ \textbf{/ Kernel Creation}} & \makecell{\textbf{Heterogeneous} \\ \textbf{Inference}} & \makecell{\textbf{Parallelization} \\ \textbf{Acceleration}} \\
    \midrule
    SoD$^2$~\citep{niu2024sod2} & \checkmark & \xmark & \checkmark & Not Discussed \\
    MikPoly~\citep{yu2024optimizing} & \checkmark & \xmark & \checkmark & Not Discussed \\
    NN-Stretch~\citep{wei2023nn} & \xmark & \xmark & \checkmark & \checkmark \\
    CoDL~\citep{jia2022codl} & \xmark & \xmark & \checkmark & \checkmark \\
    ASPEN~\citep{park2023aspen} & \xmark & \xmark & \xmark & \checkmark \\
    \textbf{Parallax (Ours)} & \checkmark & \checkmark & \checkmark & \checkmark \\
    \bottomrule
  \end{tabular}
  }
  \vspace{-15pt}
\end{table}
\begin{itemize}
    \item \textbf{Graph Analysis and Parallel Scheduling}: A non-invasive DAG analysis that identifies and schedules heterogeneous subgraph execution without modifying the model.
    \item \textbf{Branch-Aware Memory Management}: Dedicated memory arenas per branch with region-based allocation and buffer reuse to eliminate contention and minimize footprint.
    \item \textbf{Device-Adaptive Scheduling}: A resource-constrained algorithm that enforces an adaptive memory budget to maximize safe parallel CPU utilization.
\end{itemize}

\section{Related Work}
\textbf{Offline Model Compression.} A common approach to optimize DNN inference on resource-constrained devices is offline model compression. Techniques such as quantization reduce numerical precision (e.g., to INT8) to shrink model size and accelerate computation~\citep{kim2022integer,yao2020int8}, while pruning removes less important weights or channels to lower FLOPs~\citep{jiang2022model}. Operator fusion merges adjacent kernels to reduce memory access and boost throughput~\citep{niu2021dnnfusion}. Architecture-level designs like MobileNets~\citep{howard2017mobilenets} and MobileBERT~\citep{sun2020mobilebert} offer efficiency through depthwise convolutions and compact Transformer blocks. Recent innovations, such as LookupFFN~\citep{zeng2023lookupffn}, replace matrix multiplications with table lookups. Although these methods reduce latency and memory, they often require retraining or calibration data, may degrade accuracy, and often involve model tuning. Critically, they do not address runtime penalties from dynamic behavior or unsupported operators. Parallax complements these approaches by directly handling fallback scenarios at runtime without further model changes.

\textbf{Hardware Acceleration and Heterogeneous Execution.} Standard mobile frameworks (e.g., TFLite, ORT, ExecuTorch) use hardware delegation and compiler optimizations to accelerate DNN inference, but rely on static graphs. To improve acceleration, prior work explored co-execution and graph partitioning. ASPEN~\citep{park2023aspen} dynamically schedules fine-grained tiles for CPU-only parallelism. $\mu$Layer~\citep{kim2019mulayer} distributes layers across CPU and GPU based on workload and datatype performance; CoDL~\citep{jia2022codl} enables intra-operator parallelism across processors. NN-Stretch~\citep{wei2023nn} restructures sequential DNNs into parallel branches for heterogeneous cores, and BAND~\citep{jeong2022band} schedules multiple DNNs across systems. OPTiC~\citep{wang2018optic} optimizes CPU-GPU partitioning and core frequencies under thermal constraints. While effective for static models, these methods often depend on heuristic-based partitioning and require careful coordination across separate memory spaces, which can introduce complexities and overhead.

\textbf{Dynamic Operations and Memory Management.} To address static graph limitations, SoD$^2$~\citep{niu2024sod2} uses symbolic analysis to infer tensor shapes and generate optimized code paths, while MikPoly~\citep{yu2024optimizing} pre-compiles shape-specialized micro-kernels for runtime linking. Although these methods offer flexible solutions, they rely on offline analysis and dispatch generation that require regeneration as models or operators evolve. Hardware-centric solutions like Tandem Processor~\citep{ghodrati2024tandem} integrate programmable cores for irregular operations, but such designs require new silicon and are incompatible with the existing edge device ecosystem. SOTA framework memory allocators~\citep{david2021tensorflow,bai2019} minimize memory via aggressive buffer reuse but create data dependencies that block branch-level parallelism. Parallax overcomes this by assigning dedicated memory arenas per branch with efficient reuse within and across branches, minimizing contention and enabling reliable parallelism. Combined with lightweight memory estimation and adaptive scheduling, Parallax maximizes concurrency within tight memory budgets and remains adaptable across diverse models and edge platforms.

\section{Parallax}
\begin{figure}[t]
\begin{center}
\includegraphics[width=.9\textwidth]{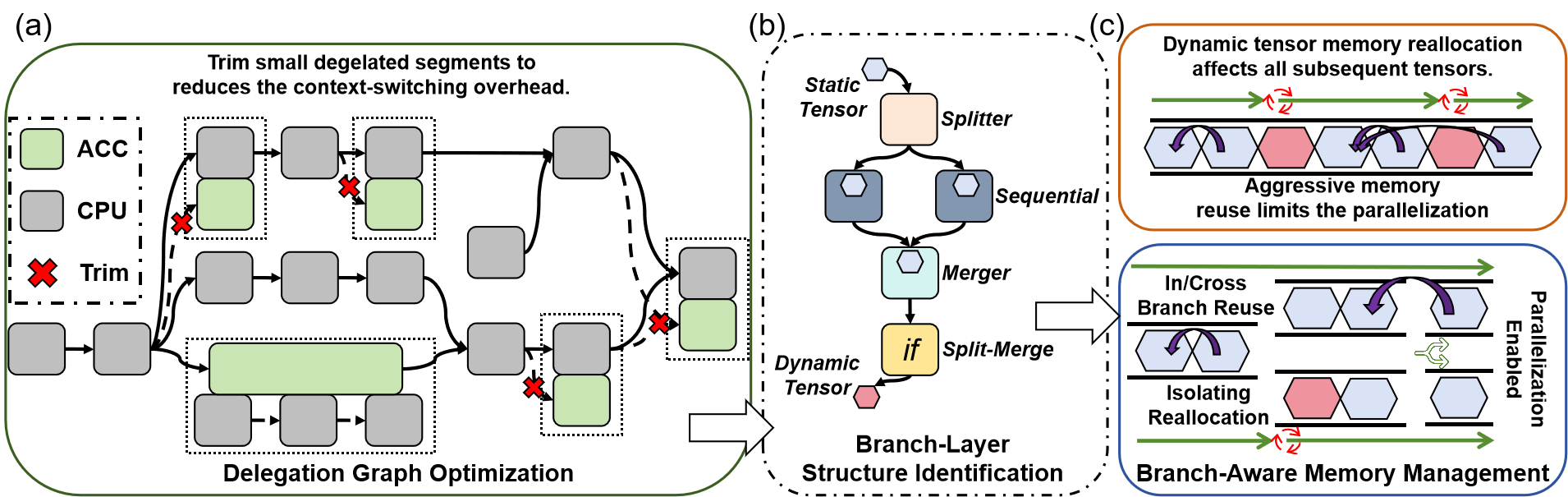}
\end{center}
\caption{Parallax overview. (a) Delegation graph optimization trims small delegated segments to reduce synchronization overhead. (b) Branch-layer structure identification classifies nodes to expose parallel execution paths. (c) Branch-aware memory management isolates branch arenas to avoid contention and safely reuse arenas, enabling efficient parallel execution under dynamic tensor shapes.}
\label{fig:system}
\end{figure}
\textbf{Problem Formulation.} Mobile DNN inference increasingly encounters dynamic graphs, creating two major challenges: \textbf{(i) Parallel execution disruption.} Mixed graphs with CPU fallbacks and delegateable segments are executed sequentially, causing repeated host-device transfers, synchronization delays and idle CPU cores. \textbf{(ii) Inefficient memory handling.} Static memory planners allocate global heaps based on fixed shapes. When shapes are resolved only at runtime (e.g., conditional branches, dynamic decoding), allocators must invalidate and reallocate large regions, adding overhead and limiting parallelism. Existing methods either overlook these problems or rely on offline preprocessing and static kernel generation, which struggle to generalize across evolving models.

Parallax addresses these challenges through a compiler-level framework. It applies graph analysis, optimized delegation and fallback handling to expose parallelizable branches and reduce synchronization overhead. The branch-aware memory allocation further minimizes unnecessary reallocations, supporting efficient execution under dynamic shapes. The design comprises three coordinated stages (Fig.~\ref{fig:system}), detailed in the following subsections.
\subsection{Graph Analysis and Partitioning}
\label{sub:graph}
As the foundation step, Parallax addresses an underexplored challenge: restructuring complex, heterogeneous DNN computation graphs to expose independent, parallelizable execution paths. Without this, fallback regions remain fragmented and tightly coupled to delegate segments, limiting parallelism and efficient memory planning. We preprocess the graph $\mathcal{G} = (\mathcal{V}, \mathcal{E})$, where $\mathcal{V}$ are operations and $\mathcal{E}$ are tensor dependencies, to (i) identify accelerator-worthy regions, (ii) expose branch-level parallelism in mixed CPU-delegate graphs, and (iii) generate per-branch workload metadata for later stages.

\textbf{Optimized Delegate Partitioning.} Identifying accelerator-worthy regions within dynamic, heterogeneous DNN graphs is both challenging and essential: naively offloading small or low-compute subgraphs causes kernel-launch and data-transfer overheads that negate any acceleration gains. Parallax applies an analytical cost model, grounded in representative mobile SoC parameters, to prune inefficient delegate candidates (Fig.~\ref{fig:system}a).

We characterize each candidate region \(S\) by (1) its operation count \(N = |V(S)|\), (2) total compute \(F = \sum_{v\in S}\mathrm{FLOPs}(v)\) MACs (Multiply–Accumulate operations) estimated using operator-level FLOPs (Appendix \ref{appendix:flop}) and (3) boundary transfer size \(B = \sum_{T\in\partial S}\mathrm{numel}(T)\times\mathrm{sizeof(dtype)}\), where \(\partial S\) is the set of boundary tensors between \(S\) and the rest of the graph, \(\mathrm{numel}(T)\) is the number of elements in tensor \(T\), and \(\mathrm{sizeof(dtype)}\) is the byte width of the tensor’s data type. A region is offloaded only if:
\[
N\ge 3,\quad
F\ge 1\times10^{9},\quad
\frac{B}{F}\le 0.1.
\]
The thresholds arise from requiring that total offload time
\[
T_{\text{offload}} = L + \frac{F}{R_{\mathrm{acc}}} + \frac{B}{B_{\mathrm{bw}}}
\]
be less than CPU execution time \(\tfrac{F}{R_{\mathrm{cpu}}}\), where \(L\) is the accelerator dispatch latency, \(R_{\mathrm{acc}}\) is the accelerator’s peak throughput (in MACs/s), and \(B_{\mathrm{bw}}\) is the peak memory bandwidth (in bytes/s). This simplifies (Appendix~\ref{appendix:cost_model}) to:
\[
F > L R_{\mathrm{acc}},\quad \frac{B}{F} < \frac{B_{\mathrm{bw}}}{R_{\mathrm{acc}}}.
\]
We adopt representative values from published hardware reports: (i)median mobile GPU dispatch latency \(L = 0.2\,\mathrm{ms}\), typical for NNAPI burst mode across devices~\citep{kim2021minimizing}; (ii) peak accelerator throughput \(R_{\mathrm{acc}} = 2.6\times10^{13}\,\mathrm{MAC/s}\), matching Qualcomm Snapdragon 8 Gen 1 specifications~\citep{qualcomm2021snapdragon8gen1}; (iii) memory bandwidth \(B_{\mathrm{bw}} = 51.2\,\mathrm{GB/s}\), reflecting LPDDR5 in 2021–2022 flagship SoCs~\citep{qualcomm2020snapdragon888}. Substituting yields \(F > 5.2\times10^9\) MACs and \(\tfrac{B}{F} < 0.002\) bytes/MAC. We relax these to \(F \ge 1\times10^9\) and \(\tfrac{B}{F} \le 0.1\) to account for device variability, kernel inefficiencies and runtime scheduling fluctuations. 

\textbf{Branch and Layer Extraction.} 
After delegate partitioning, the graph intermixes accelerator‐offload and CPU‐fallback nodes whose dependencies limit independent scheduling. 
\begin{figure}[h]
\vspace{-10pt}
\centering
\begin{minipage}[t]{0.48\textwidth}
\begin{algorithm}[H]
\caption{Node Classification and Branch Identification}
\label{alg:nc}
\begin{algorithmic}[1]
\Require Graph $\mathcal{G} = (\mathcal{V}, \mathcal{E})$
\Ensure Branch set $\mathcal{B}$
\ForAll{$v \in \mathcal{V}$}
    \State compute $(d_{in}(v), d_{out}(v))$
    \State label $v$ as Sequential / Splitter / Merger / Split-Merge
\EndFor
\State $\mathcal{B} \gets \emptyset,\ \mathcal{V}_{vis} \gets \emptyset$
\ForAll{$v \in \mathcal{V}$ if $v \notin \mathcal{V}_{vis}$ and not Merger/Split-Merge}
    \State $b \gets [\,]$
    \While{$v$ Sequential and $v \notin \mathcal{V}_{vis}$}
        \State append $v$ to $b$; mark $v$ visited; $v \gets$ successor
    \EndWhile
    \State add $b$ to $\mathcal{B}$
\EndFor
\end{algorithmic}
\end{algorithm}
\end{minipage}
\hfill
\begin{minipage}[t]{0.48\textwidth}
\begin{algorithm}[H]
\caption{Layer Construction via Topological Sort}
\label{alg:lc}
\begin{algorithmic}[1]
\Require Branches $\mathcal{B}$, dependencies $\mathcal{E}$
\Ensure Layers $\mathcal{L}$
\State compute $d[b] \gets$ in-degree for $b \in \mathcal{B}$
\State initialize empty queue $Q\gets\{b\mid d[b]=0\}$
\State initialize empty list $\mathcal{L}\gets[\,]$
\While{$Q \neq \emptyset$}
    \State $layer\gets Q$; $Q\gets\emptyset$
    \ForAll{$b \in layer$}
        \State process branch $b$; remove from $Q$
        \ForAll{$b'$ dependent on $b$}
            \State decrement $d[b']$; add $b'$ to $Q$ if $d[b']=0$
        \EndFor
    \EndFor
    \State append $layer$ to $\mathcal{L}$
\EndWhile
\end{algorithmic}
\end{algorithm}
\end{minipage}
\vspace{-2pt}
\end{figure}
Parallax decomposes the graph into a Branch-Layer structure that proceeds in two stages. First, Parallax classifies each node by connectivity:  
\emph{Sequential} (\texttt{in}=1, \texttt{out}=1),  
\emph{Splitter} (\texttt{in}=1, \texttt{out}>1),  
\emph{Merger} (\texttt{in}>1, \texttt{out}=1), or  
\emph{Split-Merge} (\texttt{in}>1, \texttt{out}>1).  
Delegate regions are treated as indivisible units, while control‐flow operators (e.g., \texttt{If}, \texttt{While}) are marked Split-Merge to ensure sequential correctness. Maximal branches are then extracted via traversal as shown in Algorithm~\ref{alg:nc}. Finally, branches are grouped into layers via a topological sort over the branch dependency map~\ref{alg:lc}. This entire process runs in $O(|\mathcal V| + |\mathcal E|)$ time and outputs $\mathcal B$ and $\mathcal L$, which enable efficient parallel scheduling and memory management in subsequent stages (Fig.~\ref{fig:system}b). (Expanded algorithm listings are provided in Appendix~\ref{app:partitioning}.)

\textbf{Further Refinement.} Parallax further refines its execution plan by enforcing minimal workloads for parallelizable layers and balancing workloads across branches within each layer. To quantify these workloads, we reuse the metrics \(N\) and \(F\). For a layer's branches to execute in parallel, each branch must satisfy:
\[
N > 2,\quad \text{and}\quad \frac{F_{\max}}{F_{\min}} \leq \beta
\]
where \(F_{\max}\) and \(F_{\min}\) denote the FLOPs of the heaviest and lightest branches, respectively. The threshold \(\beta\) was empirically determined (\(1.5\) in experiments) to balance workload effectively without incurring synchronization overhead.

\subsection{Branch-Aware Memory Management}
\label{sub:bmm}
Parallel execution of branches requires careful memory isolation to prevent contention and unsafe reuse. Parallax assigns each branch \( b_i \) a dedicated memory arena \( \mathcal{A}_i \), ensuring all tensor allocations remain within \( \mathcal{A}_i \) to avoid cross-branch conflicts (Fig.~\ref{fig:system}c).

\textbf{In-Branch Memory Reuse.}
Within each arena, Parallax uses a bump-pointer allocator with liveness analysis. When a tensor’s last use completes, its buffer is reclaimed into a free list for reuse by subsequent tensors. Buffer reuse is safe if tensor lifetimes do not overlap:
\begin{equation}
    \text{reuse}(T_j, T_k) \Longleftrightarrow \text{lifetime}(T_j) \cap \text{lifetime}(T_k) = \varnothing.
\end{equation}
This approach minimizes fragmentation and reduces peak memory usage, which is critical for resource-constrained mobile inference.

\textbf{Cross-Arena Buffer Sharing.}
Parallax also enables safe cross-arena buffer reuse. When branches \( b_i \) and \( b_j \) reside in different, non-concurrent layers, freed buffers from \( \mathcal{A}_i \) can be transferred to \( \mathcal{A}_j \). This further reduces peak memory without incurring synchronization overhead.

\textbf{Handling Dynamic Tensor Shapes.}
Dynamic operations generate tensors with unknown shapes at runtime, requiring on-the-fly memory (re)allocations. Naive handling risks synchronization delays or overwriting buffers of concurrent branches. Parallax avoids this by confining all dynamic tensor allocations and resizes to the arena \( \mathcal{A}_i \) of the originating branch. This isolation eliminates cross-branch conflicts and enables safe parallel execution under constrained memory.

\subsection{Resource-Constrained Parallel Scheduling}
\label{subsec:scheduling}
Even with balanced and memory-isolated branches, running parallel branches simultaneously could exceed the device’s memory capacity. To prevent OOM issues while maximizing available concurrency, Parallax finally employs a resource-constrained scheduling strategy.

\textbf{Branch Peak Memory Estimation.} Parallax estimates the peak memory \(M_i\) of each branch \(b_i\) in three steps. First, \emph{shape inference} computes tensor sizes from operator metadata (dimensions, data types). Second, \emph{liveness analysis} tracks each tensor’s lifetime within the branch; tensors needed downstream remain active. Finally, a linear scan over interval endpoints maintains a running total of active memory, recording the peak as \(M_i\). This sweep runs in \(O(|V|)\) time with negligible overhead, as it is fused with branch structure identification (\S\ref{sub:graph}).

\textbf{Greedy Layer Scheduling.} During runtime, Parallax continuously queries the operating system for available free memory. Then it set a safety margin of 30-50\% to be the working memory budget \(M_{\text{budget}}\). Within each layer, Parallax then selects the largest possible subset of branches whose combined estimated peak memory does not exceed this budget:
\[
\textstyle \sum_{b_i \in \text{chosen}} M_i \;\le\; M_{\text{budget}}.
\]
Branches not selected for parallel execution are run sequentially. This simple yet effective strategy avoids OOM failures while maximizing safe concurrency within the device's memory limits.

\section{Experiment}
Parallax is integrated into TensorFlow 2.18.1 via targeted modifications to \texttt{Invoke}, \texttt{InvokeImpl} and \texttt{PartitionGraphIntoIndependentNodeSubsets}, focusing on \texttt{Subgraph}, \texttt{Interpreter}, and \texttt{GraphInfo} to enable graph partitioning and branch parallelization. A custom memory planner extends \texttt{SimpleMemoryArena} for branch-aware memory management and isolated parallel execution. We also develop an evaluation app to benchmark Parallax on Android devices with Kirin, Dimensity and Google Tensor processors, measuring latency, peak memory and energy under a unified protocol.

\subsection{Experimental Setup}
\begin{table}[h]
\centering
\vspace{-10pt}
\caption{Summary of DNN models. Symbolic input dimensions represent variable input size.}
\label{tab:model_summary}
\scalebox{0.8}{
\begin{tabular}{l l l l r}
\toprule
\textbf{Model} & \textbf{Task} & \textbf{Input Shape} & \textbf{Precision} & \textbf{Params} \\
\midrule
YOLOv8n~\citep{yolov8_ultralytics2023} & Object detection & [1, 3, 640, 640] & FP32 & 3.19M \\
Whisper-Tiny~\citep{radford2023robust} & Speech recognition & [1, 3000] & INT8/FP32 & 46.51M \\
SwinV2-Tiny~\citep{liu2022swinv2} & Image classification & [1, 3, 224, 224] & FP16 & 28.60M \\
CLIP Text Encoder~\citep{radford2021clip} & Text embedding & [batch, sequence\_len] & FP32 & 63.17M \\
DistilBERT~\citep{sanh2019distilbert} & Sentiment Classification  & [batch, sequence\_len] & FP32 & 66.96M \\
\bottomrule
\end{tabular}
}
\end{table}
\textbf{Baselines \& Benchmarks.} We evaluate Parallax across 5 representative DNN models, summarized in Table~\ref{tab:model_summary}. Comparisons are made against three SOTA mobile inference frameworks: ORT, ExecuTorch and TFLite. For consistency, all models were initially implemented in PyTorch and subsequently converted to the target formats of each framework. An exception was Whisper-Tiny, where publicly available, framework-specific open-source implementations were used due to conversion difficulties. All baselines were evaluated on identical hardware and software setups.

\textbf{Testing Platforms.} Experiments used three smartphones: Google Pixel 6 (Google Tensor SoC, TPU, 8-core CPU, 2.80GHz), Huawei P30 Pro (Kirin 980 SoC, GPU, 8-core CPU, 2.60GHz), and Redmi K50 (Dimensity 8100 SoC, MDLA/DSP/GPU, 8-core CPU, 2.85GHz). Hardware offloading was enabled via Android’s Neural Networks API (NNAPI) for heterogeneous inference. Although Kirin 980 includes a Mali-G76 GPU and dual-core NPU, neither is NNAPI-accessible. In this case, we leverage the OpenCL backend in TFLite to enable heterogeneous inference.

\textbf{Performance Metrics.} We report end-to-end inference latency (averaged over 20 runs after 5 warm-up runs), peak runtime memory usage and estimated energy consumption measured with Android Studio Energy Profiler. Benchmark inputs include random images from the COCO validation set (YOLOv8n), audio samples from LibriSpeech test-clean (Whisper-Tiny), images from the ImageNet validation set (MobileNetV2, SwinV2-Tiny), and sentences randomly selected from the SST-2 dataset (CLIP Text Encoder, DistilBERT), 30 for each.
\subsection{Results \& Discussion}
Parallax leaves model weights and structure unchanged, ensuring identical outputs and accuracy to the original pretrained models. Therefore, while improving inference latency and energy, it does not affect functional performance across evaluations.
\begin{table}[t]
\centering
\caption{End‐to‐end inference latency (ms) on three devices. Each entry reports the minimum / maximum latency. "Het" denotes heterogeneous inference. "-" indicates not supported due to operator-set mismatch, lack of backend support or inability to handle dynamic input tensors without manual shape fixing (in Parallax). The best results are in \textbf{bold}.}
\label{tab:latency_all_devices}
\small
\scalebox{0.83}{
\begin{tabular}{l 
                cc  
                cc  
                cc  
                cc} 
\toprule
\multirow{2}{*}{\textbf{Model}}
  & \multicolumn{2}{c}{\textbf{ORT}} 
  & \multicolumn{2}{c}{\textbf{ExecuTorch}} 
  & \multicolumn{2}{c}{\textbf{TFLite}} 
  & \multicolumn{2}{c}{\textbf{Parallax}} \\
\cmidrule(lr){2-3} \cmidrule(lr){4-5} \cmidrule(lr){6-7} \cmidrule(lr){8-9}
  & CPU & Het & CPU & Het & CPU & Het & CPU & Het \\
\midrule
\multicolumn{9}{l}{\textbf{Google Pixel 6}} \\
YOLOv8n            & 83 / 974 & - & 77 / 933 & - & 85 / 1355 & - & 63 / 794 & \textbf{54} / \textbf{743} \\
Whisper-Tiny       & 397 / 2114 & 486 / 3142 & 421 / 2376 & - & 543 / 2076 & - & 350 / 1780 & \textbf{323} / \textbf{1706} \\
SwinV2-Tiny        & 82 / 87 & 1323 / 1726 & 84 / 93 & - & 96 / 108 & 1107 / 1994 & \textbf{64} / 83 & 69 / \textbf{79} \\
CLIP Text Encoder  & 16 / 43 & 15 / 44  & 17 / 56 & - & 13 / 37 & - & \textbf{11} / \textbf{29} & - \\
DistilBERT         & \textbf{14} / 54 & 19 / \textbf{48} & 15 / 51 & - & 27 / 57 & - & 17 / 49 & - \\
\midrule
\multicolumn{9}{l}{\textbf{Huawei P30 Pro}} \\
YOLOv8n            & 136 / 1327 & - & 138 / 1475 & - & 127 / 1463 & - & 101 / 1016 & \textbf{91} / \textbf{957} \\
Whisper-Tiny       & 433 / 3298 & - & 487 / 2954 & - & 412 / 2976 & - & \textbf{340} / \textbf{2239} & - \\
SwinV2-Tiny        & 128 / 134 & - & 131 / 143 & - & 136 / 148 & 987 / 1753 & 109 / 126 & \textbf{107} / \textbf{122} \\
CLIP Text Encoder  & 18 / 68 & - & 19 / 72 & - & 17 / 72 & - & \textbf{15} / \textbf{57} & - \\
DistilBERT         & 19 / 73 & - & 17 / 83 & - & 25 / 63 & - & \textbf{14} / \textbf{61} & - \\
\midrule
\multicolumn{9}{l}{\textbf{Redmi K50}} \\
YOLOv8n            & 98 / 1032 & - & 117 / 1215 & - & 124 / 1273 & - & 91 / 912 & \textbf{84} / \textbf{881} \\
Whisper-Tiny       & 501 / 3326 & 634 / 3793 & 496 / 3182 & - & 499 / 2980 & - & \textbf{317} / \textbf{2010} & - \\
SwinV2-Tiny        & 86 / 94 & - & 91 / 98 & - & 83 / 89  & 1046 / 1935 & 54 / 62 & \textbf{53} / \textbf{59} \\
CLIP Text Encoder  & \textbf{16} / 46 & 14 / 51 & 17 / 51 & - & 21 / 49 & - & \textbf{16} / \textbf{43} & - \\
DistilBERT         & \textbf{15} / 55 & 38 / 77 & 16 / 61 & - & 21 / 66  & - & 17 / \textbf{47} & - \\
\bottomrule
\end{tabular}
}
\end{table}

\textbf{Latency Performance.} Table~\ref{tab:latency_all_devices} reports Parallax latency versus baselines. In CPU-only execution, Parallax consistently outperforms others, with 15–31\% reductions on large-input models such as YOLOv8n ($3\times640\times640$) and Whisper-Tiny (30-second audio). For text encoders with 16–77 tokens (CLIP, DistilBERT), Parallax shows comparable latency, with minor overheads (e.g., DistilBERT on Pixel 6) from branch scheduling.

In heterogeneous execution, ORT remains the strongest baseline for dynamic inputs (e.g., CLIP, DistilBERT), as ExecuTorch lacks NNAPI support and TFLite reverts to CPU for dynamic operators. Parallax’s fine-grained subgraph control and branch-isolation memory enable partial offloading to accelerators, yielding 9–46\% latency gains over ORT and 20–45\% over TFLite. Gains are most notable on complex models like Whisper-Tiny and SwinV2-Tiny, where conventional frameworks suffer from context-switching overhead due to fragmented delegation. These results demonstrate that Parallax effectively leverages parallel branch execution and selective offloading to mitigate performance penalties for dynamic and fragmented models in heterogeneous edge environments.
\begin{table}[t]
\centering
\vspace{-5pt}
\caption{Peak runtime memory usage (MB), including model static memory and dynamic temporary activations or tensor allocation. Lowest peak memory results are in  \textbf{bold} across devices.}
\label{tab:memory_all_devices}
\small
\scalebox{0.75}{
\begin{tabular}{l 
                cccc  
                cccc  
                cccc} 
\toprule
\multirow{2}{*}{\textbf{Model}}
  & \multicolumn{4}{c}{\textbf{Pixel 6}} 
  & \multicolumn{4}{c}{\textbf{P30 Pro}} 
  & \multicolumn{4}{c}{\textbf{Redmi K50}} \\
\cmidrule(lr){2-5}\cmidrule(lr){6-9}\cmidrule(lr){10-13}
  & ORT & ET & TFLite & Parallax 
  & ORT & ET & TFLite & Parallax 
  & ORT & ET & TFLite & Parallax \\
\midrule
YOLOv8n           & 24.9 & 25.1 & \textbf{24.5} & 26.6   & 29.6 & 28.3 & 29.7 & 32.5   & 34.2 & - & 26.7 & 33.0 \\
Whisper-Tiny      & 50.1 & 48.2 & 47.8 & 63.6   & 52.4 & 49.6 & \textbf{45.3} & 72.9 & 51.8 & - & 49.6 & 71.2 \\
SwinV2-Tiny       & 86.7 & 84.9 & \textbf{82.3} & 107.6   & 102.3 & 95.7 & 96.4 & {132.0}  & 97.3 & 99.1 & 102.4 & 124.8 \\
CLIP Text Encoder & 137.2 & 140.7 & 134.6 & 184.4   & 143.5 & 139.2 & 142.0 & 191.2   & 135.9 & 139.0 & \textbf{133.7} & 174.3 \\
DistilBERT        & 206.7 & 203.2 & \textbf{197.6} & 233.5   & 215.9 & 211.1 & 207.4 & 234.7  & 230.4 & 227.3 & 229.0 & 265.2 \\
\bottomrule
\end{tabular}
}
\end{table}

\begin{table}[t]
\centering
\caption{Peak memory footprint (MB) of tensor arena allocations across frameworks. "Naive" denotes no liveness reuse, and every tensor gets separate memory.}
\label{tab:memory_arena}
\small
\begin{tabular}{lccccc}
\toprule
\textbf{Model} & \textbf{ORT} & \textbf{ExecuTorch} & \textbf{TFLite} & \textbf{TFLite (Naive)} & \textbf{Parallax} \\
\midrule
YOLOv8n & 69.98 & 72.45 & 74.28 & 203.62 & 85.82 \\
Whisper-Tiny & 28.94 & 35.12 & 31.67 & 90.15 & 43.20 \\
SwinV2-Tiny & 17.22 & 12.87 & 31.67 & 38.40 & 46.75 \\
CLIP Text Encoder & 3.21 & 4.09 & 3.42 & 7.69 & 4.97 \\
DistilBERT & 4.67 & 5.18 & 4.02 & 9.41 & 7.42 \\
\bottomrule
\end{tabular}
\end{table}
\textbf{Runtime Memory Analysis.} Table~\ref{tab:memory_all_devices} shows Parallax peak memory versus baselines across devices. Concurrent branch execution moderately increases memory but stays within device limits due to resource-constrained scheduling (\S\ref{subsec:scheduling}). The largest rise is on Whisper-Tiny (72.9\,MB vs. 45.3\,MB, +60.9\% on P30 Pro), with smaller increases for YOLOv8n (33.0\,MB vs. 26.7\,MB, +23.6\% on Redmi K50) and SwinV2-Tiny (124.8\,MB vs. 102.4\,MB, +21.9\%). On average, memory rises 26.5\%, more pronounced for large fragmented models (Whisper, SwinV2), but negligible for smaller workloads (CLIP, DistilBERT).

Table~\ref{tab:memory_arena} further quantifies Parallax’s branch-aware allocator overhead (\S\ref{sub:bmm}). Its arena footprint averages 46.3\% larger than TFLite and 37.7\% larger than ORT, reflecting our choice of branch isolation over aggressive reuse. Compared to a naïve planner (one buffer per tensor), Parallax reduces arena size by 43.2\% (like YOLOv8n: 85.8\,MB vs. 203.6\,MB, –57.8\%). This demonstrates that Parallax achieves effective intra-branch memory reuse while enabling parallel execution across branches. We argue this memory overhead represents a worthwhile trade-off, as it supports the substantial latency improvements required for real-time edge inference and ensures memory remains predictable and bounded under the parallel execution model. This also preserves deployment flexibility by eliminating the need for model refactoring or custom kernel modifications.

\begin{figure}[t]
  \centering
  \begin{minipage}[t]{0.48\textwidth}
    \centering
    \includegraphics[width=.9\textwidth]{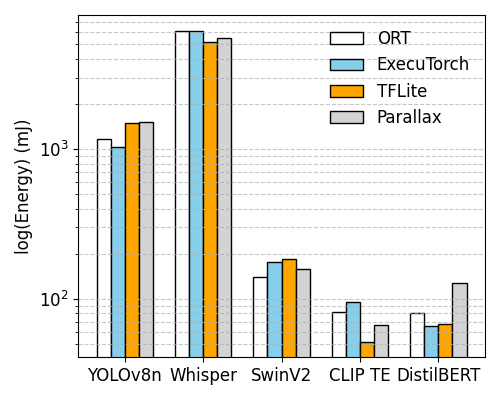}
    \caption{Energy performance on Google Pixel 6. Results are for CPU-only inference, which supports the majority of models.}
    \label{fig:energy}
  \end{minipage}
  \hfill
  \begin{minipage}[t]{0.48\textwidth}
    \centering
    \includegraphics[width=.9\textwidth]{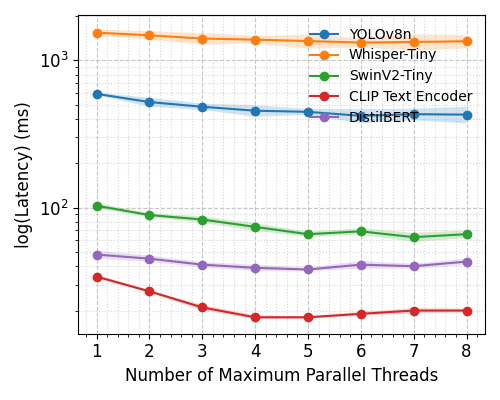}
    \caption{Maximum parallel threads vs. latency for Parallax (Pixel 6 CPU-only inference).}
    \label{fig:threads_ablation}
  \end{minipage}
\end{figure}
\textbf{Energy Consumption.} Figure~\ref{fig:energy} shows energy costs across frameworks on Google Pixel 6. Parallax achieves energy savings through reduced inference time. On Whisper-Tiny, energy decreases by 10.0\% compared to ORT and 10.7\% compared to ExecuTorch. Similarly, for the CLIP Text Encoder, Parallax lowers energy by 18.3\% and 30.0\% compared to ORT and ExecuTorch, respectively. For SwinV2-Tiny, Parallax energy rises 13.3\% over ORT but improves 14.1\% over TFLite. For YOLOv8n and DistilBERT, parallel execution overhead increases energy up to 47.0\% and 92.2\% vs. ExecuTorch, but still remains within practical mobile inference. Note that during all tests, device battery temperature fluctuation remained modest, ranging from \SI{27.9}{\degreeCelsius} to \SI{33.4}{\degreeCelsius}.

Although TFLite generally achieves the lowest peak memory and better energy in some models, it runs 1.2–1.8× slower than Parallax. ORT  provides a marginal improvement than Parallax for latency in dynamic models, but Parallax improves efficiency via branch-level parallel scheduling, offering a compelling alternative for energy-constrained real-time inference.
\subsection{Ablation Study}
\begin{table}[t]
\centering
\caption{Layer-wise latencies and branch counts for Whisper (CPU) vs.\ SwinV2-Tiny (CPU+TPU).}
\label{tab:layer_latency_sidebyside}
\small
\scalebox{0.9}{
\begin{tabular}{c r r c @{\quad} c r r c}
\toprule
\multicolumn{4}{c}{\textbf{Whisper (CPU)}} 
  & 
\multicolumn{4}{c}{\textbf{SwinV2-Tiny (CPU+TPU)}} \\
\cmidrule(lr){1-4} \cmidrule(lr){5-8}
\textbf{Layer ID} & \textbf{TFLite (ms)} & \textbf{Parallax (ms)} & \textbf{BR.} 
  & 
\textbf{Layer ID} & \textbf{TFLite (ms)} & \textbf{Parallax (ms)} & \textbf{BR.} \\
\midrule
3   & 37.26 & 27.59 & 3    & 2   &  4.32 &  2.97    & 4        \\
7   &  6.37 &  4.28 & 6    & 10  &  1.59 &  1.66    & 1        \\
8   & 12.40 & 12.64 & 1    & 37  &  7.45 &  5.09    & 6        \\
24  & 45.37 & 39.74 & 8    & 53  & 13.46 &  7.61    & 4 (1D+3) \\
30  &  2.35 &  2.27 & 1    & 103 &  2.33 &  2.41    & 1 (D)    \\
\bottomrule
\end{tabular}
}
\end{table}
\textbf{Layer-Level Latency Analysis.} We profiled Whisper (CPU) and SwinV2-Tiny (CPU+TPU) on Pixel 6 using our graph analysis method (\S\ref{sub:graph}). Table~\ref{tab:layer_latency_sidebyside} shows selected layers with TFLite vs. Parallax latencies and branch counts (BR). On multi-branch layers, Parallax achieves up to 32.8\% improvement on Whisper layer 7 (6 branches) and 43.5\% on SwinV2 layer 53 (1 TPU + 3 CPU branches), with averages of 23.7\% and 35.5\%. Single-branch layers, including delegated ones, show minor overheads ($\le4.4\%$) from thread setup and coordination. Overall, Parallax accelerates parallelizable segments while imposing negligible penalties on sequential layers.

\begin{table}[t]
\centering
\caption{Graph Structure and parallelism performance on Pixel 6. "Pre" represents pre-delegation graphs i.e. original graphs; "Post" means delegated graphs; "Parallax" means graphs optimized by Parallax.}
\label{tab:parallelism_detailed}
\small
\scalebox{0.85}{
\begin{tabular}{l rrr rrr rrr rrr}
\toprule
\multirow{2}{*}{\textbf{Model}}
  & \multicolumn{3}{c}{\textbf{Nodes}}
  & \multicolumn{3}{c}{\textbf{Layers}}
  & \multicolumn{3}{c}{\textbf{Par-Layers}}
  & \multicolumn{3}{c}{\textbf{Max-Branches}} \\
\cmidrule(lr){2-4} \cmidrule(lr){5-7} \cmidrule(lr){8-10} \cmidrule(lr){11-12}
  & Pre & Post & Parallax 
  & Pre & Post & Parallax 
  & Pre & Post & Parallax 
  & Pre & Post & Parallax \\
\midrule
YOLOv8n            & 480 &   5 &   4 & 120 & 6 & 4 & 69 &  2 &  1 & 6 & 2 & 2 \\
Whisper-Tiny       & 627 & 202 & 367 & 75 & 184 & 57 & 19 & 10 & 24 & 8 & 5 & 8 \\
SwinV2-Tiny        &1108 & 356 & 674 & 151 & 270 & 143 & 92 & 28 & 52 & 8 & 8 & 8 \\
CLIP Text Encoder  & 635 & 216 & 446 & 90 & 173 & 74 & 53 & 19 & 49 & 4 & 3 & 4 \\
DistilBERT         & 353 & 123 & 211 & 53 & 115 & 49 & 28 & 23 & 17 & 4 & 2 & 4 \\
\bottomrule
\end{tabular}
}
\end{table}
\textbf{Graph Optimization Effect.} Table~\ref{tab:parallelism_detailed} summarizes how Parallax transforms each model’s graph (\S\ref{sub:graph}). After initial operator delegation ("Post"), node counts drop sharply (e.g., Whisper-Tiny: 627 to 202), but graphs fragment into fine-grained layers (Whisper-Tiny: 75 to 184; SwinV2-Tiny: 151 to 270). Parallax applies graph partitioning to fallback small delegate regions, yielding more compact structures. It also exposes greater parallelism: parallelizable layers increase from 10 to 24 (Whisper-Tiny) and from 28 to 52 (SwinV2-Tiny), while maximum concurrent branches recover from 5 to 8 in both. These results show Parallax simplifies fragmented graphs while improving scheduling flexibility and execution efficiency.

\textbf{Thread Parallelism Sensitivity.} Figure~\ref{fig:threads_ablation} shows the effect of varying Parallax’s maximum parallel threads on Google Pixel 6. Latency decreases as threads increase, especially between 1 and 4, where parallelism is best utilized. Beyond 4 threads, improvements fluctuate based on model branch count and workload. For smaller models (CLIP, DistilBERT), excessive threads introduce management overhead, offsetting gains.

We set Parallax’s maximum thread count to 6 in experiments. For fairness, other frameworks were also set to 6 where applicable. We observed similar trends when increasing their number of threads: using all available cores sometimes degraded performance significantly due to scheduling overhead (e.g., in TFLite multi-threaded inference). This safe setting provided stable and comparable results across all platforms.

\section{Conclusion and Future Work}
We presented Parallax, a framework that accelerates heterogeneous mobile inference through non-invasive graph analysis, fine-grained subgraph control, branch-aware memory management, and resource-constrained parallel scheduling. On five DNNs across three smartphones, Parallax delivers 15–31\% CPU speedups, 9–46\% heterogeneous-mode latency reductions, and up to 30\% energy savings. Despite a moderate 26.5\% memory overhead, Parallax balances real-time performance and deployment simplicity, requiring no model refactoring or custom kernels. Ablation studies confirm the benefits of graph partitioning, branch isolation, and thread scaling.

Despite these advantages, Parallax has three main limitations that we plan to address in future work: \textbf{(i) Partial Dynamic‐Flow Support.} Parallax currently provides only fine-grained subgraph delegation for dynamic tensors and control-flow constructs, leaving unsupported operators on the CPU. Enabling full accelerator support for arbitrary dynamic shapes and flows remains an open challenge for mobile inference; \textbf{(ii) Energy Overhead.} Our current evaluation focuses on latency and memory management without optimizing for energy efficiency. Some models show higher net energy consumption due to branch scheduling and increased power draw. Integrating energy-efficient scheduling and graph optimization is a key area for future improvement; \textbf{(iii) Model Conversion and Compatibility.} Real-world model deployment often encounters precision mismatches, unsupported operators or backend-specific challenges. Parallax lacks the discussion about these aspects, but we are extending our graph-analysis framework with automated conversion and delegation-assignment tools to support a broader range of models with minimal manual intervention.

By addressing these challenges, we aim to further enhance Parallax’s applicability, efficiency and deployment scalability across diverse mobile inference scenarios.

\bibliography{neurips_2025}

\newpage
\appendix
\section{FLOP Estimation and Delegation Threshold Justification}
\label{appendix:flop}
\subsection{Operator-Level FLOP Estimation}
To support Parallax’s delegation pruning (\S\ref{sub:graph}), we estimate the compute intensity \( F \) of candidate regions using operator-level FLOP counts. We categorize TFLite operators into coarse-grained groups, each with a simplified FLOP estimator:

\begin{table}[h]
\centering
\caption{Operator Classes and FLOP Estimators}
\vspace{0.5em}
\begin{tabular}{l|l|l}
\textbf{Op Class} & \textbf{Examples} & \textbf{FLOPs per Node} \\
\hline
Conv2D / Depthwise & Conv2D, DepthwiseConv2D & 
$2 \cdot C_\text{in} \cdot H_\text{out} \cdot W_\text{out} \cdot K_h \cdot K_w \cdot C_\text{out}$ \\
MatMul / Dense & FullyConnected, MatMul & $2 \cdot M \cdot N \cdot K$ \\
Elementwise & Add, Mul, ReLU, Sub & $\text{output\_size}$ \\
Pooling / Reduce & AvgPool, MaxPool, Mean, Sum & $H_\text{out} \cdot W_\text{out} \cdot K_h \cdot K_w$ \\
Misc. / Other & Reshape, Slice, Transpose & $0$ (or $0.5 \cdot \text{output\_size}$ optionally) \\
\end{tabular}
\label{tab:flop-estimation}
\end{table}

This coverage accounts for the majority of compute in modern CNNs and Transformers. Unrecognized or non-compute-heavy ops are either treated as 0-FLOP or assigned a small constant workload.

\section{Delegation Cost Model Derivation}
\label{appendix:cost_model}
This section provides the full derivation and numerical justification for the delegate–pruning thresholds used in §3.1.

\subsection{Offload vs.\ CPU Execution Time}
Consider a candidate region \(S\) with
\[
F = \sum_{v\in S}\text{FLOPs}(v)
\quad\text{and}\quad
B = \sum_{T\in\partial S}\text{numel}(T)\times\text{sizeof(dtype)}.
\]
Then:
\[
T_{\text{offload}}
= L \;+\; \frac{F}{R_{\mathrm{acc}}} \;+\; \frac{B}{B_{\mathrm{bw}}}
\]
is the total time to dispatch, compute on the accelerator and transfer boundary tensors back.
Meanwhile, running the same region on the CPU takes
\[
T_{\text{CPU}}
= \frac{F}{R_{\mathrm{cpu}}}
\]
where \(R_{\mathrm{cpu}}\) is the CPU’s MAC/s rate.

We choose to offload only when
\[
T_{\text{offload}} < T_{\text{CPU}}
\quad\Longleftrightarrow\quad
L + \frac{F}{R_{\mathrm{acc}}} + \frac{B}{B_{\mathrm{bw}}}
< \frac{F}{R_{\mathrm{cpu}}}.
\]

\subsection{Simplified Bounds}
Since \(R_{\mathrm{cpu}}\!\ll\!R_{\mathrm{acc}}\), we can decompose the above condition into two practical sub-conditions:

\paragraph{Compute-bound condition.}  
Neglecting memory transfer (\(B/B_{\mathrm{bw}}\approx0\)) and using \(R_{\mathrm{cpu}}\ll R_{\mathrm{acc}}\), the inequality reduces to
\[
L + \frac{F}{R_{\mathrm{acc}}} \;<\; \frac{F}{R_{\mathrm{cpu}}}
\;\approx\; \frac{F}{R_{\mathrm{cpu}}}
\;\Longrightarrow\;
F \;>\; L\,R_{\mathrm{cpu}}.
\]
Thus, the region’s total MACs \(F\) must exceed the number of MACs a CPU can perform during the dispatch latency \(L\).
\begin{figure}[h]
   \centering
   \begin{minipage}[t]{0.48\textwidth}
    \begin{algorithm}[H]
    \caption{Node Classification and Branch Identification}
    \label{alg:node_classification}
    \begin{algorithmic}[1]
    \State \textbf{Input}: Computation graph $\mathcal{G} = (\mathcal{V}, \mathcal{E})$
    \State \textbf{Output}: Set of branches $\mathcal{B}$
    \For{each node $v \in \mathcal{V}$}
        \State Compute in-degree $d_{in}(v)$ and out-degree $d_{out}(v)$
        \If{$d_{in}(v) = 1$ and $d_{out}(v) = 1$}
            \State Label $v$ as \texttt{Sequential}
        \ElsIf{$d_{in}(v) = 1$ and $d_{out}(v) > 1$}
            \State Label $v$ as \texttt{Splitter}
        \ElsIf{$d_{in}(v) > 1$ and $d_{out}(v) = 1$}
            \State Label $v$ as \texttt{Merger}
        \ElsIf{$d_{in}(v) > 1$ and $d_{out}(v) > 1$}
            \State Label $v$ as \texttt{Split-Merge}
        \EndIf
    \EndFor
    \State Initialize empty set of branches $\mathcal{B}$
    \State Initialize visited set $\mathcal{V}_{visited} \gets \emptyset$
    \For{each node $v \in \mathcal{V}$}
        \If{$v \notin \mathcal{V}_{visited}$ and $v$ is not labeled as \texttt{Merger} or \texttt{Split-Merge}}
            \State Initialize new branch $b \gets [\,]$
            \While{$v$ is labeled as \texttt{Sequential} and $v \notin \mathcal{V}_{visited}$}
                \State Append $v$ to $b$
                \State Add $v$ to $\mathcal{V}_{visited}$
                \State $v \gets$ successor of $v$
            \EndWhile
            \State Add branch $b$ to $\mathcal{B}$
        \EndIf
    \EndFor
    \end{algorithmic}
    \end{algorithm}
   \end{minipage}
   \hfill
   \begin{minipage}[t]{0.48\textwidth}
    \begin{algorithm}[H]
    \caption{Layer Construction via Topological Sorting}
    \label{alg:layer_construction}
    \begin{algorithmic}[1]
    \State \textbf{Input}: Set of branches $\mathcal{B}$, dependencies $\mathcal{E}$
    \State \textbf{Output}: Ordered list of layers $\mathcal{L}$
    \State Initialize in-degree map $d: \mathcal{B} \rightarrow \mathbb{N}$
    \For{each branch $b \in \mathcal{B}$}
        \State $d[b] \gets$ number of incoming dependencies to $b$
    \EndFor
    \State Initialize queue $Q \gets$ branches with $d[b] = 0$
    \State Initialize empty list of layers $\mathcal{L} \gets [\,]$
    \While{$Q$ is not empty}
        \State Initialize empty list $layer \gets [\,]$
        \For{each branch $b \in Q$}
            \State Append $b$ to $layer$
            \For{each branch $b'$ dependent on $b$}
                \State Decrement $d[b']$ by 1
                \If{$d[b'] = 0$}
                    \State Add $b'$ to $Q$
                \EndIf
            \EndFor
        \EndFor
        \State Append $layer$ to $\mathcal{L}$
    \EndWhile
    \end{algorithmic}
    \end{algorithm}
   \end{minipage}
\end{figure}
\paragraph{Memory-bound condition.}  
Neglecting compute time (\(F/R_{\mathrm{acc}}\approx0\)), offloading helps only if
\[
L + \frac{B}{B_{\mathrm{bw}}} \;<\; \frac{F}{R_{\mathrm{cpu}}}.
\]
However, more commonly one asks whether the accelerator itself would be memory-bound: comparing compute to transfer on the accelerator gives
\[
\frac{B}{B_{\mathrm{bw}}} \;<\; \frac{F}{R_{\mathrm{acc}}}
\quad\Longleftrightarrow\quad
\frac{B}{F} \;<\; \frac{B_{\mathrm{bw}}}{R_{\mathrm{acc}}}.
\]
This ensures the accelerator spends more time computing than waiting on memory.

\subsection{Numerical Substitution}
Using the representative SoC parameters from \S\ref{sub:graph}:
\[
L = 0.2\ \mathrm{ms},\quad
R_{\mathrm{cpu}} \approx 1 \times 10^{9}\ \mathrm{MAC/s},\quad
R_{\mathrm{acc}} = 2.6\times10^{13}\ \mathrm{MAC/s},\quad
B_{\mathrm{bw}} = 51.2\times10^{9}\ \mathrm{B/s},
\]
we obtain:
\[
L\,R_{\mathrm{cpu}} = 0.2\times10^{-3}\times10^{9} = 2\times10^{5}\ \text{MACs},
\quad
\frac{B_{\mathrm{bw}}}{R_{\mathrm{acc}}}
= \frac{51.2\times10^{9}}{2.6\times10^{13}}
\approx 0.002\ \text{bytes/MAC}.
\]
To provide a safety margin for driver overhead, runtime jitter and variability across devices, we conservatively enforce:
\[
F \;\ge\; 1\times10^{9}
\quad\text{and}\quad
\frac{B}{F} \;\le\; 0.1\ \text{bytes/MAC},
\]
and since each operator typically costs on the order of \(10^8\)–\(10^9\) MACs, we also require
\(\;N = |V(S)|\ge 3\)
to ensure meaningful per-region compute.

\section{Graph Partitioning and Scheduling Algorithms}
\label{app:partitioning}
To support the layer-branch abstraction (\S\ref{sub:graph}), Parallax employs two algorithms:
\begin{itemize}
    \item \textbf{Node Classification and Branch Identification:} This algorithm classifies each node in the computation graph $\mathcal{G} = (\mathcal{V}, \mathcal{E})$ based on its in-degree and out-degree, and identifies branches as maximal linear sequences of nodes without internal splits or merges, shown in Algorithm \ref{alg:node_classification}.
    \item \textbf{Layer Construction via Topological Sorting:} This algorithm constructs layers by performing a topological sort on the identified branches, ensuring that all dependencies are respected and that branches within the same layer can be executed in parallel, shown in Algorithm \ref{alg:layer_construction}.
\end{itemize}

\end{document}